Chapter 5

**Two-Level Systems and the Tunneling Model: A Critical View**

Clare C. Yu and Hervé M. Carruzzo

*Department of Physics and Astronomy,
University of California, Irvine
Irvine, CA 92697
cyu@uci.edu*

The thermal and acoustic properties displayed by a wide variety of glasses at low temperatures are well described by the model of tunneling two level systems (TLS). We review the standard TLS model as well as developments that have occurred since the earlier work. In spite of half a century of effort and impressive successes, mysteries still remain. We discuss some of these longstanding puzzles about glasses at low temperatures and speculate about future directions.

## 1. Introduction

The goal of this article is to raise questions about the properties of glasses at low temperatures.[1,2] So let us start with this one, "Why are the thermal properties of glasses different from those of crystals at low temperatures?" Crystals have a phonon (Debye) specific heat and a thermal conductivity that both go as $T^3$ at low temperatures, independent of the crystal structure. This is due to phonons whose wavelengths λ are much larger than the lattice constant *a*; as a result the phonons are insensitive to differences in crystal structures. This insensitivity to crystal structure could lead one to predict that the specific heat and thermal conductivity of all insulators should behave similarly at low temperatures. But this prediction would be wrong.

Amorphous materials below 1 K have a specific heat that is linear in temperature and a thermal conductivity that is quadratic in temperature (see Chapter 1, Figures 1 and 2).[3] Between 3 and 10 K, the thermal conductivity displays a plateau followed by a rise at higher temperatures.[3] In the temperature range of the plateau, the specific heat displays a bump when $C/T^3$ is plotted versus $T$ on log-log plot.[3] This bump is associated with the so-called 'boson peak'. If *C/T* vs. *T* is plotted, there is a steep rise by 2 to 3 orders of magnitude between 1 and 20 K (Figure 1).

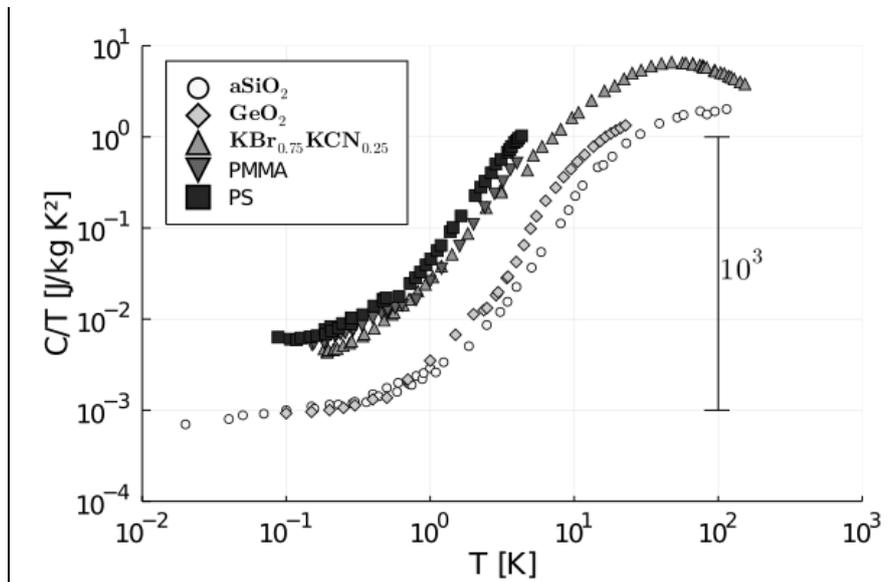

Figure 1. *C/T* vs. *T* for amorphous $SiO_2$ (a$SiO_2$) (Refs. 3, 4), $GeO_2$ (Refs. 3, 5), $KBr_{0.75}CN_{0.25}$ (Ref. 6), PMMA (Ref. 7,8,9), and PS (Ref. 8,9) shows a steep rise by 2 to 3 orders of magnitude between 3-10 K.

It is remarkable that the amorphous structure, rather than chemical composition, determines the thermal, acoustic and dielectric properties of glassy materials. What is it about the disordered structure that produces such universal behavior?

### 1.1. *Standard model of two level systems*

To explain the behavior below 1 K, the standard model of tunneling two level systems (TLS) was proposed by Anderson, Halperin, and Varma[10] and independently by Phillips.[11] The basic idea is that there are dynamic defects consisting of an atom or group of atoms that can sit more or less equally well in either of two energy minima of a double well potential. At low temperatures the atom or group of atoms tunnels back and forth between the two configurations represented by the two minima. The energy difference between the two minima is given by the energy asymmetry ε. The potential barrier height is $V$ while $d$ is the distance between the minima. In the right-well-left-well basis, the Hamiltonian is given by:

$$H = \frac{1}{2}\begin{pmatrix} \varepsilon & \Delta_o \\ \Delta_o & -\varepsilon \end{pmatrix} \qquad (1)$$

The tunneling matrix element, $\Delta_o$, is given by the WKB formula

$$\Delta_o = \omega_o e^{-W} \qquad (2)$$

where

$$W \sim \frac{\sqrt{2MV}\,d}{\hbar} \qquad (3)$$

$M$ is the mass of the tunneling entity, and the attempt frequency $\omega_o$ is on the order of the Debye frequency since atomic motion is involved in the tunneling. The two level systems that tunnel fast enough to be seen experimentally must have low barrier heights ($V \leq 0.1$ eV). In order to conserve energy, a phonon must be emitted or absorbed when a transition is made between energy levels given by the energy eigenvalues:

$$E_\pm = \pm\frac{1}{2}\sqrt{\varepsilon^2 + \Delta_o^{\,2}} \qquad (4)$$

This implies that the energy splitting $E = E_+ - E_- \leq k_B T$. A typical estimate for the density of two level systems is 100 ppm which comes from integrating the specific heat of $SiO_2$ up to 10 K. The TLS model assumes that the distribution $N(\varepsilon)$ of asymmetry energies is flat and the distribution $P(W)$ of the WKB exponent is flat.[10, 11] Even though $N(\varepsilon)$ probably varies on the scale of 0.1 eV, it is reasonable to expect it to be flat for energies of order $k_B T$ where $T \sim 1$ K.[10] On the other hand, the distribution for the WKB exponent $W$ does not have such a natural explanation besides the fact that it leads to a flat distribution of energy splittings. Regardless, the assumption is then that $P(\varepsilon, W) = \overline{P}$ where $\overline{P}$ is a constant. Changing variables from $W$ to $\Delta_o$ yields

$$P(\varepsilon, \Delta_o) = \frac{\overline{P}}{\Delta_o} \tag{5}$$

As a result, the distribution $n(E)$ of energy splittings $E$ in Eq. (4) is a constant $n_o$. It is straightforward to show (see Chapter 2) that this uniform density of states produces a linear specific heat using an approach similar to that used to calculate the linear heat capacity of an electron gas.

To calculate the thermal conductivity, we note that up to about 50 to 60 K, Zaitlin and Anderson showed that heat is transported by phonons in insulating glasses.[12] The thermal conductivity is given by

$$\kappa(T) = \frac{1}{3} \int_0^{\omega_D} C_{ph}(T,\omega) v \ell(T,\omega) d\omega \tag{6}$$

where $\omega_D$ is the Debye frequency, $C_{ph}$ is the phonon specific heat which goes as $T^3$, $v$ is the sound velocity, and $\ell(T,\omega)$ is the phonon mean free path. The phonon mean free path is limited by resonant scattering from TLS in which a phonon with energy $\hbar\omega$ is absorbed by a TLS with energy splitting $E \simeq \hbar\omega$. This requires that phonons couple to the strain field as described by the following term in the Hamiltonian:

$$H_{TLS-Ph} = \gamma e \sigma_z \tag{7}$$

where $\gamma$ is the coupling constant, $e$ is the strain field and $\sigma_z$ is the z-component Pauli matrix. Calculating the resonant scattering rate yields $\ell \sim \omega^{-1} \sim T^{-1}$, which, when combined with $C_{ph} \sim T^3$, produces $\kappa(T) \sim T^2$. In fitting the thermal conductivity data to Eq. (6), Freeman and Anderson[13] noted the remarkable result that $\ell \sim 150\lambda$ below about 200 GHz and for temperatures around and below 1 K for a wide variety glasses.

### 1.2. *Crossover Region*

As we mentioned earlier, between 3-10 K there is a plateau in the thermal conductivity and a bump in $C/T^3$ that corresponds to the boson peak. In fitting the thermal conductivity to Eq. (6) to find the frequency dependence of the mean free path $\ell$ below about 10 K, Freeman and Anderson[13] found that $\ell \sim 150\lambda$ at low frequencies, drops precipitously around 200 GHz, and then $\ell \sim \lambda$ at higher frequencies (see Fig. 4 in Chapter 1). One can regard the plateau as a crossover from a long mean free path to a regime with a short mean free path. One way to account for these observations is to ascribe them to a steep rise at an energy $E_o$ in the density of states of dynamic defects that scatter phonons.[12, 14, 15] This is consistent with the bump in $C/T^3$, i.e., the rise in $C/T$. Fitting the specific heat data results in $E_o$ between 10 and 40 K.[15] This is consistent with the rise in the density of states observed in Raman scattering[16] and neutron scattering.[17, 18] ($E_o/k_B$ being greater than the temperature of the bump in $C/T^3$ is due to excitations being able to contribute to the specific heat at temperatures well below their characteristic energy.) Yu and Freeman[15] used this rise in the density of states of phonon scatterers to fit the specific heat and thermal conductivity of various glasses up to roughly 100 K. In order to fit both the plateau in the thermal conductivity and the bump in $C/T^3$ with the same density of states, they had to introduce strong (Rayleigh) scattering at low frequencies ($\hbar\omega < E_o$) but not at high frequencies ($\hbar\omega > E_o$).

Zaitlin and Anderson were unable to find physically realistic values for the coefficient of Rayleigh scattering due to density fluctuations.[12] However, it is possible that Rayleigh scattering could arise from other sources, e.g., fluctuations in local force constants,[19] though they did not consider this. Zaitlin and Anderson were able to fit the thermal conductivity without invoking Rayleigh scattering by adding a term quadratic in the energy to the density of states of the phonon scatterers and cutting off the steep drop in the phonon mean free path at high frequencies with a minimum mean free path of order a few angtroms.[12, 14] However, unlike Yu and Freeman,[15] they did not try to do a detailed fit to the specific heat with this same density of states.

These findings raise several questions: First what are the excess excitations responsible for the boson peak? This question is still not settled, though it has been discussed extensively in the literature[20] and is reviewed in chapters 7-11 of this book. Second, what picks out the energy scale of $E_o$ for the rise in the density of states? It is much less than the other energy scales in the problem, e.g., the Debye energy is of order 300 K and electronic energies are of order thousands of degrees. Third, what is the source of the strong (Rayleigh) scattering and why does the strong (Rayleigh) scattering appear to stop at high frequencies? If Rayleigh scattering of phonons were allowed to occur at all frequencies, the thermal conductivity would not rise above the plateau. However, it is quite likely that above the plateau, where the phonon mean free path is of order the wavelength, ballistic phonons are no longer operating. Rather, phonon diffusion or hopping of quasi-localized vibrations may be transporting heat.[21, 22] These are questions worth pondering but let us return to the subject of two level systems.

## 2. Beyond the Standard Model of Two Level Systems

### 2.1. *Interacting Two Level Systems*

Since TLS couple to the strain field, TLS can interact with each other by exchanging phonons.[23, 24] Using second order perturbation theory, one can show that this interaction has a dipolar form that goes as $g/r^3$ where $g \sim \gamma^2/\rho v^2$, $r$ is the distance between the two interacting TLS, and $\rho$ is the mass density.[23, 24] Yu and Leggett argued that this interaction should not be neglected.[25] To see why, let us estimate $g/r^3$ as follows. If we take $\gamma \sim 1\,\text{eV}$ (a typical experimentally deduced value), then $g \sim 5\times 10^4$ K-Å$^3$. For a TLS concentration of 100 ppm, $r^3 \sim 10^4$ Å$^3$, assuming 1 atom/Å$^3$. Putting this all together yields $g/r^3 \sim 5$ K $>>$ 0.05 K $\sim$ 1 GHz, the typical TLS energy splitting $\Delta E$ seen in ultrasound experiments.[26, 27] Thus, interactions between TLS cannot be ignored. Indeed, some experiments can only be explained by including interactions.

Historically, TLS-TLS interactions were largely ignored in the standard TLS model except to account for spectral diffusion.[28] For example, in acoustic hole burning experiments,[29] a high power acoustic pulse is resonantly absorbed by those TLS whose energy splitting matches the frequency of the pulse, leading to an increase in the population of the upper energy level and to a decrease in the attenuation. A probe pulse reveals a "hole" in the resonant attenuation centered at the frequency of the high power pulse. The width of the hole is much wider than one would expect from the typical decay rate of excited TLS.[29] To understand this,[28, 29] we note that the energy splitting of any given TLS will fluctuate adiabatically due to the fluctuating strain field produced by nearby fluctuating TLS that are undergoing thermal excitations and relaxations back to their ground state. Thus the broadening of the width of the hole in the attenuation is due to fluctuations in the energy splittings of the excited TLS. While the explanation of spectral diffusion experiments requires TLS-TLS interactions, other types of experiments, e.g., specific heat and thermal conductivity, did not, and so, such interactions were largely ignored.

Interactions between TLS were not widely incorporated into the description of glasses at low temperatures until the work of Yu and Leggett[25] and other theoretical efforts,[30-42] together with subsequent nonequilibrium dielectric experiments from the Osheroff group.[43, 44] In those experiments, an ac capacitance bridge was used to study the dielectric response of glasses such as $SiO_2$, $SiO_x$, and polymers. Slow dc sweeps showed a minimum in the real part of the ac dielectric susceptibility χ' at the electric field $E_o$ in which the sample was cooled. In the absence of an applied bias voltage ($E_o = 0$), the hole in χ' is centered at zero bias. Upon switching the dc electric field to a new value $E_{new}$ different from $E_o$, χ' increased abruptly and then decayed logarithmically in time. If the electric field is held fixed at $E_{new}$, a second minimum develops at that field as revealed by subsequent dc field sweeps. The abrupt increase of χ' upon the sudden application of $E_{new}$ is due to χ' jumping out of the bottom of the hole, followed by the logarithmic relaxation in time corresponding to digging a new hole.

An analysis of these experiments by Carruzzo, Grannan, and Yu[45] showed that while many features could be explained with the model of non-interacting TLS, the frequency dependence of the slope, *dC/d(ln t)*, of the logarithmic relaxation of the capacitance *C* required interacting TLS. In particular, interactions lead to the formation of clusters of TLS, with large slow clusters contributing to the dielectric response only at low frequencies. Interactions also provide a natural explanation for the hole in χ' which reflects the hole in *P(h)*, the distribution of local fields *h*, produced by neighboring TLS. (These fields could be local strain and/or electric fields.) Stability arguments[46-48] require that there be a hole in *P(h)* as *h* → 0. To see this, consider a spin glass with long-range interactions. Suppose there is a ground state spin configuration with *P(h=0)* ≠ 0. This implies that there are spins that have zero local field and that, therefore, can flip without any change in energy. But if these spins flip, this would change the local fields of other spins and so, some of them would flip. This in turn would cause still other spins to flip, leading to a proliferation of flipping spins. This avalanche means that the supposed ground state is not stable. So in order to have a stable ground state, *P(h)* must go to zero as *h* → 0. This hole in *P(h)* is sometimes referred to as the dipole gap. Carruzzo *et al.* used Monte Carlo simulations of an Ising spin glass to show that upon abrupt application of a dc field, the susceptibility jumps up and then relaxes logarithmically, with interactions leading to frequency dependence of the slope of the logarithmic relaxation.[45] An Ising spin glass could be considered as the limit of densely interacting TLS, though Carruzzo *et al.* did not specify the distance between nearest neighbor Ising spins.

Burin[44, 49] also analyzed the Osheroff experiments. Like Carruzzo *et al.*,[45] he found that these experiments could not be explained by non-interacting TLS and that interactions between TLS lead to a dipole gap in the density of states. He considered the limit of dilute interacting TLS and argued that slowly relaxing pairs of TLS lead to slow, logarithmic relaxation following an abrupt change in the dc field.

### 2.2. *Universally small phonon scattering*

Not only do glasses exhibit qualitatively universal behavior, measurements also reveal quantitatively universal values. Phonon scattering is an example. We have already mentioned that fitting the thermal conductivity below 1 K leads to the ratio $\ell/\lambda \approx 150$, where $\ell$ is the phonon mean free path and $\lambda$ is the phonon wavelength.[13] This is equivalent to the universal value found for the inverse Q factor, sometimes called internal friction, which is a measure of ultrasonic attenuation. Recall that if one mounts a sample on an oscillator, e.g., a double paddle oscillator, and measures the amplitude as a function of frequency, then $Q^{-1}$ is the ratio of the resonant linewidth to the resonant frequency $f_o$, i.e., $Q^{-1} = \Delta f / f_o$. $Q^{-1}$ is proportional to $\lambda / \ell$:

$$Q^{-1} = \frac{\lambda}{4\pi\ell} \tag{8}$$

Measurements on a wide variety of glasses find that $Q^{-1}$ lies in the range between $10^{-4}$ and $10^{-3}$, with a typical value of $3 \times 10^{-4}$. The universality of this value has been a long-standing puzzle. Why is phonon scattering so insensitive to the composition and structure of different amorphous materials?

Within the context of the standard TLS model, phonons scatter from TLS and the measured quantities can be expressed in terms of *C*, a dimensionless scattering rate:

$$C = \frac{\bar{P}\gamma^2}{\rho v^2} \tag{9}$$

where $\gamma$ is the strength of the TLS-phonon coupling, $\rho$ is the mass density of the glass, and *v* is the velocity of sound given by

$$\frac{1}{v^3} = \frac{1}{3}\sum_s \frac{1}{v_s^3} \tag{10}$$

where $v_s$ is the sound velocity for polarization *s*. In Eq. (9), $\bar{P}$ is the same $\bar{P}$ found in Eq. (5) and corresponds to the density of states of TLS that tunnel fast enough to scatter phonons, a small subset of all the TLS that contribute to the specific heat. The TLS density of states, $n_o$, that enters into the specific heat is typically about ten times larger than $\bar{P}$.

Various measured quantities can be expressed in terms of *C*. For example,

$$Q^{-1} = \frac{\pi}{2}C \tag{11}$$

The change in sound velocity in the relaxation regime is given by

$$\frac{\Delta v}{v} = -\frac{1}{2} C \ln\left(\frac{T}{T_o}\right) \qquad (12)$$

where $T_o$ is an arbitrary reference temperature. In the relaxation regime, phonons perturb the TLS energy-level separation, and as a result, the level population must readjust to a new Boltzmann equilibrium. In terms of $C$, the ratio of the mean free path to the wavelength can be expressed as

$$\frac{\ell}{\lambda} = \frac{1}{2\pi^2 C} \qquad (13)$$

Measurements of these quantities yield values of $C$ between $2 \times 10^{-4}$ and $5 \times 10^{-4}$ for a wide variety of amorphous materials. This universal value of $C$ is quite surprising given the large variation from glass to glass of the parameters entering into $C$ in Eq. (9).[50] It is also odd that a dimensionless number should result in a number so far from unity. Fits to experimental measurements on insulating glasses yield TLS-phonon couplings $\gamma$ on the order of 1 eV,[51] a puzzling energy scale since it does not match any other in the problem.

Yu and Leggett[25] (YL) made the first attempt to explain the universal value of $C$ using a mean field approach in which the interactions between TLS determined the density of states, resulting in

$$\bar{P} \sim \frac{\rho v^2}{\gamma^2} \qquad (14)$$

Plugging this in Eq. (9) yields $C \sim 1$ which is universal but of the wrong order of magnitude, though it is consistent with $\ell \sim \lambda$ found at temperatures above the plateau in the thermal conductivity.

Subsequently, a number of other authors have proposed a variety of rather complicated approaches to achieve the right order of magnitude for $C$.[52-56] Burin and Kagan[52] obtain a renormalized value of $\bar{P}$ in Eq. (5) by considering a dilute system of TLS with two-body interactions, $J_{ij} = g_{ij}/r_{ij}^3$, between TLS at sites $i$ and $j$. The coupling $g_{ij}$ has a typical value of $g_o$. As the temperature $T$ decreases, the physics is dominated by increasingly large TLS clusters with constituent pairs that satisfy $T < J_{ij}$. Using their renormalized value of $\bar{P}$, Burin and Kagan find that $C = \bar{P} g_o$ is of order $10^{-3}$. Vural and Leggett[53] used a block scaling renormalization group approach in which TLS were replaced by blocks represented by stress tensors that interacted with one another via elastic dipolar interactions. In going to longer length scales by combining smaller blocks to form larger blocks, they find that $Q^{-1}$ vanishes in the thermodynamic limit. They rescue this by postulating an ansatz for the form of $Q^{-1}$. Lubchenko and Wolynes[54] related their theory of the glass transition as a random first-order (phase) transition (RFOT) to two level systems at low temperatures. In their RFOT theory, a cluster of cooperatively rearranging particles forms a droplet with a typical volume of $\xi^3$. At low temperatures, the lowest two energy levels of a droplet constitute a two-level system. Lubchenko and Wolynes[54] argue that

$$C \sim \left(\frac{\xi}{a}\right)^3 \sim \frac{1}{200} \qquad (15)$$

where $a$ the lattice constant. This is certainly the right order of magnitude, but identifying which particles are in a droplet and which are not, in a molecular dynamics simulation, say, is a challenging proposition.

Schechter and Stamp[56] proposed two types of TLS. One set has weak interactions with phonons due to inversion symmetry, and hence, weak elastic interactions between TLS, while the other type has strong interactions with phonons, and hence, strong elastic interactions between TLS. The ratio of the TLS-phonon coupling constants associated with these two types of interactions naturally gives rise to the right order of magnitude for the universally small value of the ultrasonic attenuation as well as to the small energy scale of 3 K.

Parshin[41] invoked the soft potential model to try to explain the universal value of $C$. The soft potential model postulates that the excitations in glasses at low temperature are due to quasilocal soft anharmonic oscillators described by quartic polynomials.[57-59] Some of these are double well potentials that correspond to TLS while others are anharmonic single potential wells. By making assumptions and estimates of various parameters that enter the model, e.g., those that govern the distribution of the coefficients of the fourth order polynomials describing the anhamonic potentials, Parshin was able to arrive at a small value for *C* consistent with experiment. Somewhat later, Parshin, Schober and Gurevich[55] considered localized harmonic oscillators coupled to high frequency modes such as phonons. This coupling leads to a vibrational instability that produces quasilocal harmonic vibrations that populate the boson peak as well as double well potentials that correspond to tunneling two level systems. In their model, *C* is a sensitive function of the coupling of the localized harmonic oscillators to the high frequency modes. They point out that only a small fraction of TLS are able to tunnel fast enough to contribute to experimental measurements, and it is this fraction that determines the small value of *C*. Note that while the aforementioned models have been able to obtain the right order of magnitude for *C*, their use of a variety of assumptions and estimates have precluded their ability to make quantitative predictions of the value of *C* for specific glasses.

To avoid the cancellation that Yu and Leggett (YL)[25] found, Carruzzo and Yu[60] treated $\bar{P}$ and $\gamma^2/\rho v^2$ separately in Eq. (8). They revisited the standard model of TLS and exploited aspects that had not been previously appreciated. First, they pointed out that the Pauli matrix $\sigma_z$ in Eq. (7) means that phonons actually couple to the *difference* between the elastic dipole moments in the two wells of the TLS double well potential. Since the elastic dipole moment in each well should be randomly oriented, the difference should also be random, varying from TLS to TLS. This leads to a broad distribution of couplings $\gamma$, in contrast to the standard TLS model where $\gamma$ is taken to be a constant of order 1 eV. They find that a flat distribution for $\gamma^2$ with a maximum value $\gamma_{max} \sim \rho v^2 v_o \sim 5$ eV which is much larger than 1 eV. ($v_o$ is the volume of a chemical formula unit.) Like YL, Carruzzo and Yu[60] assumed that TLS-TLS interactions dictate the density of states $\bar{P}$, but since $\gamma_{max}$ is so large, the resulting value of $\bar{P}$ is much lower, and more realistic, than what YL found. Carruzzo and Yu also pointed out that the TLS coupling to phonons leads to an exponential reduction of the tunneling matrix element due to phonon overlap between the two wells; this is a kind of polaron effect.[61] The idea is that when a TLS tunnels to one well, the neighboring atoms around it relax to accommodate it, making it harder to tunnel to the other well. Unlike the standard TLS model where tunneling depends on the WKB exponent,[10, 11] Carruzzo and Yu found integrating out the high frequency phonons produces a tunneling matrix element that depends exponentially on the coupling:

$$\Delta_o = \Delta_{o,\max} e^{-\gamma^2/\gamma_o^2} \qquad (16)$$

where the maximum tunneling amplitude $\Delta_{o,\max} \sim 10$ K and $\gamma_o$ depends on material parameters:

$$\gamma_o = \frac{\sqrt{2}}{3}\sqrt{\rho v^2 v_o \hbar \omega_D} \qquad (17)$$

where $\omega_D$ is the Debye frequency. Eq. (16) implies that the broad distribution $\gamma^2$ will lead to a wide range of tunneling amplitudes. The exponential reduction of the tunneling rate with increasing $\gamma$ means that TLS with large TLS-phonon couplings will have very small tunneling rates and will not be seen, e.g., in internal friction measurements. So only those TLS that tunnel fast enough to respond on an experimental time scale will contribute to measurements. This leads to an experimentally determined effective coupling $\gamma_{\text{eff}} \sim 1$ eV. (This energy scale is set by $\gamma_o$ given in Eq. (17). Plugging typical values into Eq. (17) yields $\gamma_o \sim 0.5$ eV.) Using $\gamma_{\text{eff}} \sim 1$ eV in $\gamma^2/\rho v^2$ in Eq. (9), together with the revised value of $\bar{P}$, leads to the correct order of magnitude for $C$. Since their theory expressed various quantities in terms of material parameters, Carruzzo and Yu were able to make detailed comparisons of their theory with experimental values of $C$ for a variety of glasses as shown in Table 1. Overall the comparisons were very good with a few exceptions, e.g., LAT, which may be due to the inability to properly estimate $v_o$, the volume of a formula unit.

Table 1: $C_{th}$ for dielectric glasses are from Ref. [60], except polystyrene (PS). Data from Ref. [51] and references therein. $C_{exp} = (C_L + 2C_T)/3$ except for PS where $C_{exp} = C_L$. $C_L$ and $C_T$ are the longitudinal and transverse components of *C*, respectively. Note that in Ref. [60], $C_{exp} = C_T$. Experimentally, $C_T \sim C_L$.[51]

| Glass | $\rho$ [kg/m$^3$] | $v$ [m/s] | $v_o$ [Å$^3$] | $T_D$ [K] | $C_{exp} \cdot 10^4$ | $C_{th} \cdot 10^4$ |
|---|---|---|---|---|---|---|
| **aSiO$_2$** | 2200 | 4163 | 45.3 | 348 | 3.0 | 2.9 |
| **BK7** | 2510 | 4195 | 41.8 | 360 | 3.1 | 2.5 |
| **SF4** | 4780 | 2481 | 40.7 | 215 | 2.6 | 0.9 |
| **SF57** | 5510 | 2327 | 55.2 | 182 | 2.7 | 0.9 |
| **SF59** | 6260 | 2131 | 40.2 | 185 | 2.6 | 1.0 |
| **V52** | 4800 | 2511 | 61.1 | 190 | 4.6 | 0.8 |
| **BALNA** | 4280 | 2569 | 39.9 | 224 | 4.5 | 1.2 |
| **LAT** | 5250 | 3105 | 68.2 | 226 | 3.7 | 0.3 |
| **Zn glass** | 4240 | 2580 | 45.9 | 215 | 3.4 | 2.0 |
| **PS** | 1050 | 1670 | 167 | 90 | 3.6 | 2.4 |
| **PMMA** | 1180 | 1762 | 138.4 | 101 | 3.1 | 2.9 |

### 2.3. *Exceptions to the Rule*

While the values of the internal friction of most amorphous materials at low temperatures lie in the range of $10^{-4}$ to $10^{-3}$, there are exceptions. For example, amorphous silicon films with 1 at. % H produce $Q^{-1}$ over 200 times smaller than the usual glassy range[62] as shown in the following Chapter. This may be the result of stress introduced by the hydrogen impurities. This possibility is consistent the fact that $Q^{-1}$ in high-stress silicon nitride ($Si_3N_4$) thin films, which show no long-range order in x-ray diffraction or TEM images, is two to three orders of magnitude lower than the universal range.[63] These two experiments imply that stress can greatly reduce dissipation, perhaps by increasing the height of TLS tunnel barriers, or by reducing the TLS-phonon coupling.[64] A more prosaic reason is dissipation dilution in which the stress produces a higher resonant frequency $f_o$ and thus a lower value of $Q^{-1} = \Delta f/f_o$.[64] More recently, the Hellman group has shown that the substrate temperature during the growth of thin films of amorphous silicon can determine the amount of dissipation, with a higher growth temperature producing a denser film and a lower value of $Q^{-1}$ that is presumably due to fewer TLS.[65] Specific heat measurements support this and find that denser amorphous silicon films have substantially fewer TLS.[66] Queen *et al*. suggested that the TLS reside on microvoids which are more numerous in lower density films. Another example is the ultrastable glass, indomethacin. Thin films of indomethacin have virtually no linear term in the specific heat below 1 K, indicating an absence of TLS.[67] Recent simulations do, in fact, find that as the stability of the glass is increased, the density of TLS decreases.[68]

### 3. Mysteries about glasses at low temperatures: Open questions for the future

While there has been much progress on the low temperature properties of glasses over the past 50 years, mysteries still abound. TLS are ubiquitous but we still do not know what they are microscopically except in a few cases such as $KBr_xKCN_{(1-x)}$ in which the TLS are cyanide molecules.[69, 70] It is also not understood why TLS are found in all sorts of materials which have such a wide variety of structures and chemical compositions. Why are there so many active excitations below 1 K and what picks out the 1 K energy scale? One way to answer these questions is with computer simulations but this has been difficult because the low density of TLS (~ 100 ppm) requires systems with a prohibitively large number of atoms in order to obtain good statistics.[71] However, recently, better algorithms and computers have allowed simulations of large systems that find TLS with energy splittings smaller than 1 K.[68] Most of these TLS are very localized and only involve a few atoms, though some involve collective excitations of a couple hundred atoms.[68]

Are there really only two levels? Presumably there can be more energy levels as proposed in the soft potential model[57-59] and the model of localized oscillators coupled to phonons.[55] However, below 1 K, it makes sense to just consider the lowest energy levels. The only condition is that they not be evenly spaced in energy as required by experimental observations of saturation of the ultrasonic attenuation[27, 72] and phonon echoes.[73, 74] Saturation refers to the attenuation at a given frequency falling to zero at a given frequency as the acoustic intensity increases. The idea is that only a finite number of slowly relaxing TLS can resonantly absorb the sound. As the number of phonons exceeds the number of these TLS, the excess ultrasound will pass through the sample without being attenuated, i.e., the attenuation "saturates" (even though it is really the TLS that are saturated). Phonon echoes are the ultrasonic analog of spin echoes in nuclear magnetic resonance (NMR). Two external ultrasonic phonon pulses separated by a time $\tau$ are followed at time $2\tau$ by a spontaneous pulse, i.e., an echo. As Yu and Leggett[25] pointed out, while this implies that there are long-lived states (lifetime > $2\tau$) with time-reversal, it does not necessarily imply *two*-level systems since systems with $S \geq \frac{1}{2}$ can also produce phonon echoes.

Leggett has also pointed out that there is no direct experimental evidence that anything is actually tunneling.[25] However, it is difficult to think of another physical mechanism that produces long relaxation times at low temperatures. Evidence for long lifetimes of excited TLS come from experiments such as phonon echoes[73, 74] and the slow time dependence of the specific heat [75, 76] which is predicted to increase logarithmically with time.[77, 78]

The assumptions of the standard model of TLS lead to a flat density of states that are consistent with a linear specific heat and a quadratic thermal conductivity. However, experimentally, the temperature dependence of the specific heat is slightly superlinear[4] and the thermal conductivity is slightly subquadratic.[3, 25] This implies a mild increase with increasing energy in the density of states of the excitations contributing to the specific heat and scattering phonons that are conducting heat. A phenomenological way to describe this would be to invoke a small energy dependent correction factor, e.g., a logarithmic correction of the form[25]

$$n(E) \sim \frac{n_o}{\ln(E_o / E)} \quad (18)$$

Such a density of states would also resolve the discrepancy between the ultrasonic attenuation and the shift in the sound velocity.[27] The TLS model predicts that the ultrasonic attenuation (inverse phonon mean free path) due to resonant scattering is given by:

$$\ell^{-1}(\omega, T) = A_\ell \omega \tanh\left(\frac{\hbar \omega}{2 k_B T}\right) \quad (19)$$

where $A_\ell$ is a constant. Performing a Kramers-Kronig transformation of the sound attenuation in Eq. (19) yields the shift in the low frequency sound velocity:

$$\frac{\Delta v}{v} = \frac{2 v A_v}{\pi} \ln\left(\frac{T}{T_o}\right) \quad (20)$$

where $A_v$ is a constant and $T_o$ is a reference temperature. Eqns. (19)-(20) are consistent with experiment.[1, 2, 27] However, the TLS model predicts that $A_v = A_\ell$ while experiment finds that $A_v$ is about 40% larger than $A_\ell$.[27] One way to resolve this discrepancy is with a density of TLS states that increases slightly with increasing energy.[25, 27] However, this *ad hoc* resolution to the superlinear specific heat, subquadratic thermal conductivity and discrepancy between the ultrasonic attenuation and the shift in the velocity of sound begs the question as to the underlying reason for such an energy dependence in the density of states. We should point out that an alternative explanation for the superlinear temperature dependence of the specific heat is a time dependent specific heat that increases slowly with time.[77, 79]

Another puzzle involves the shift in the sound velocity.[2, 80] At low temperatures where resonant scattering of phonons from TLS dominates, the standard TLS model predicts that the shift in the sound velocity increases logarithmically in temperature and follows Eq. (20) which we can rewrite in terms of $C$ as

$$\frac{\Delta v}{v} = C \ln\left(\frac{T}{T_o}\right) \tag{21}$$

where $T_o$ is a reference temperature. Eq. (21) holds in the regime where $\omega \tau_{1m} \gg 1$. Here $\tau_{1m}$ is the minimum time for a TLS to relax by emitting a phonon; it corresponds to the case of a symmetric TLS with zero asymmetry energy. At higher temperatures where $\omega \tau_{1m} \ll 1$, relaxation processes can no longer be ignored and produce a velocity shift that decreases logarithmically with temperature as

$$\frac{\Delta v}{v} = -\frac{3}{2} C \ln\left(\frac{T}{T_o}\right) \tag{21}$$

When this is combined with the resonant contribution in Eq. (21), the result is:

$$\left.\frac{\Delta v}{v}\right|_{tot} = \left.\frac{\Delta v}{v}\right|_{res} + \left.\frac{\Delta v}{v}\right|_{rel} = -\frac{1}{2} C \ln\left(\frac{T}{T_o}\right) \tag{22}$$

which we also stated in Eq. (12). So the predicted ratio of the slopes between the low and high temperature velocity shifts is 2: (-1). While the observed velocity shifts are logarithmic in temperature with a positive slope at low temperature and a negative slope at high temperature, the experimentally observed ratio of the slopes is 1: (-1).[80, 81] The reason for this discrepancy with the TLS model is not understood, though the interactions between TLS may be a factor.[82]

So far we have not distinguished between longitudinal and transverse phonons, though the longitudinal and transverse sound velocities differ, as do the longitudinal and transverse TLS-phonon couplings. One puzzle is that for any given amorphous material,

$$C_L \sim C_T \sim C \tag{23}$$

where $C$ is given by Eq. (9),

$$C_s = \frac{\bar{P}\gamma_s^2}{\rho v_s^2} \tag{24}$$

and the subscript $s$ denotes $L$ (longitudinal) or $T$ (transverse).[51] There is no good explanation for why the longitudinal and transverse values of $C_s$ should be roughly the same. However, this observation leads to[51]

$$\frac{v_L^2}{v_T^2} \sim \frac{\gamma_L^2}{\gamma_T^2} \sim 2.5 \tag{25}$$

The observed ratio of the sound velocities is not so surprising since elasticity theory in an isotropic, homogeneous medium yields[53]

$$\frac{v_L^2}{v_T^2} = 3 \tag{26}$$

However, the equivalence of $C_L$ and $C_T$ as well as the equivalence of the sound velocity ratio to the ratio of TLS-phonon couplings in Eq. (25) is puzzling.

In short, there are still fundamental gaps in our understanding of glasses at low temperatures despite the impressive record of the model of two level systems (and its modifications) in explaining the observed universalities in the thermal and acoustic properties of amorphous solids.

We started this chapter by stating that naively, one would expect glasses and crystals to behave similarly at low temperatures. So why are glasses and crystals so different in their thermal and acoustic properties at low temperatures? The answer lies in the types of low energy excitations. Long length scales are typically associated with low energies while short length scales are associated with high energies. Thus, in both crystals and glasses, long wavelength phonons have low energies. However, in amorphous materials, there are also low energy excitations at short length scales, namely tunneling two level systems.[68] This may be why Vural and Leggett's renormalization group attempt to relate universal low energy behavior to long length scales has struggled to gain traction:[53] the unique features of glasses are rooted in excitations localized at short length scales.

## 4. Future Outlook

Our understanding of the low temperature properties of glasses rests on a phenomenological model. The key ingredients are tunneling and randomness in the parameters. Experiments that probe overall response of glassy systems have uncovered discrepancies with the standard TLS model that we discussed in this chapter. These experiments measure quantities averaged over the TLSs in the sample. The underlying physics is described by a model which is akin to a black box. Researchers in the field keep tweaking the black box with the hope of resolving this or that issue. This results in more complex models with diminishing explanatory power. More direct access to the behavior of individual TLS is needed. In this respect, we would be remiss not to mention recent qubit spectroscopy experiments that probe individual TLS and their interactions with strain and electric fields as well as interactions between TLS.[83-87] By probing individual TLS, these experiments will help constrain the ingredients entering the TLS model and its derivatives.

More generally, phenomenological models are a great first step in making sense of the data and posing questions. However, we would like to understand how this phenomenology emerges from broader physical considerations. There are a few fairly obvious clues as to what those considerations should be. The first clue is that the TLS model works for so many different amorphous materials, or, stated differently, that such a wide variety of glassy materials exhibit TLS phenomenology. This tells us that the amorphous nature of the system dictates the low energy excitations, much as the crystalline nature of other systems dictates their low energy excitations (phonons). The second clue comes from materials like indomethacin.[67] These ultrastable materials, while glassy, deviate markedly from the standard behavior.[67] Thus, not all amorphous structures are equal. This implies that there is at least one additional parameter needed to characterize amorphicity. Hopefully, a single set of concepts will emerge to provide a characterization that depends only on the state, e.g., structure, of the glass. A promising candidate is glass stability,[88] but other concepts may be useful, e.g., fragility.[89] The third clue is provided by the renormalization group approach of Vural and Leggett.[53] Rather than postulate the existence of TLSs, they tried to show that TLS phenomenology emerges from one ingredient common to amorphous systems, the presence of non-elastic stresses. A key assumption in their approach is the idea that the low energy excitations of tiny blocks of glass would, when assembled into bigger and bigger blocks, generate the observed behavior, leading to the conclusion that low energy physics is all you need. While appealing from a universality standpoint, the limited success of their approach leads us to conclude that the starting point must be at higher energy, a conclusion supported by the fact that the non-elastic stresses originate at the energy scale of the glass transition. Taken together, these clues tell us that we need to start from the processes happening at the glass transition itself. The two energy regimes are intimately connected. Understanding the low temperature physics of amorphous materials will have to start at the glass transition.

**Acknowledgements**

This research was supported by the U. S. National Science Foundation (NSF) through the University of Wisconsin Materials Research Science and Engineering Center (DMR-1720415).